\documentclass[reprint,twocolumn,pre,
superscriptaddress]{revtex4-1}

\usepackage{amsmath,amssymb}
\usepackage{graphicx}
 \usepackage{color}
\usepackage{siunitx}
\usepackage{changes}
\usepackage{bm}

\colorlet{Changes@Color}{red}

\newcommand{\ii}{\textrm{i}}
\newcommand{\ee}{\textrm{e}}

\newcommand{\dd}{\textrm{d}}

\newcommand{\inc}{\textrm{in}}
\newcommand{\sca}{\textrm{sc}}

\renewcommand{\vec}[1]{\bm{#1}}

\expandafter\ifx\csname package@font\endcsname\relax\else
 \expandafter\expandafter
 \expandafter\usepackage
 \expandafter\expandafter
 \expandafter{\csname package@font\endcsname}%
\fi

\begin{document}
\title
{{Acoustic spin transfer to a subwavelength spheroidal particle}}
\author{Jos\'e H. Lopes}
\affiliation{Grupo de F\'isica da Mat\'eria Condensada, N\'ucleo de Ci\^encias Exatas,
	Universidade Federal de Alagoas,
	Arapiraca, AL 57309-005, Brazil}
\author{Everton B. Lima}
\affiliation{Physical Acoustics Group,
	Instituto de F\'isica,
	Universidade Federal de Alagoas, 
	Macei\'o, AL 57072-970, Brazil}
\author{Jos\'e~P. Le\~ao-Neto}
\affiliation{Campus Arapiraca/Unidade de Ensino Penedo, Universidade Federal de Alagoas, Penedo, Alagoas
	57200-000, Brazil}
\author{Glauber T. Silva}
\email{gtomaz@fis.ufal.br}
\affiliation{Physical Acoustics Group,
Instituto de F\'isica,
Universidade Federal de Alagoas, 
Macei\'o, AL 57072-970, Brazil}

\date{\today}

\begin{abstract}
We  demonstrate that 
the acoustic spin of a first-order Bessel beam can be transferred to  a subwavelength (prolate) spheroidal particle
at the beam axis
in a viscous fluid.
The induced radiation torque is proportional to the acoustic spin, which scales with the beam energy density.
The analysis of the particle rotational dynamics in a Stokes' flow regime reveals 
that its angular velocity varies linearly with the acoustic spin.
Asymptotic expressions of the radiation torque and angular velocity are obtained for a quasispherical and infinitely thin particle.
Excellent agreement is found between the theoretical results of radiation torque and finite element simulations.
The induced particle spin is predicted and analyzed using the typical parameter values of
the acoustical vortex tweezer and levitation devices.
We discuss how the beam energy density and fluid viscosity can be assessed by measuring the induced spin of the particle.
\end{abstract}
\maketitle

\section{Introduction}
The spin angular momentum is a universal feature present in different contexts of nature.
In classical electromagnetic waves and photons, the spin is caused by the circular polarization of electric and magnetic fields~\cite{Ohanian1986}.
The electron spin can be regarded as due to a circulating flow of energy in the Dirac wave field~\cite{Belinfante1939}. 
More recently, the spin of acoustic beams
was proposed and measured as a circulation of the fluid velocity field~\cite{Shi2019}.
Subsequently,
the spin and orbital angular momenta were theoretically analyzed in monochromatic acoustic wave fields in a homogeneous
medium~\cite{Bliokh2019}.
Before these studies, it was noticed that the longitudinal spin, in which the axis of rotation is parallel to the propagation direction of an acoustic Bessel beam, could induce the acoustic radiation torque on a subwavelength absorbing spherical particle~\cite{Silva2014}.

The acoustic radiation torque is the time-averaged rate of change
of the angular momentum caused by an acoustic wave on an object~\cite{Zhang2011c}.
This subject was extensively studied for  spherical particles in Refs.~\cite{Zhang2011a,Silva2012,Mitri2012a,Zhang2013,Mitri2016,Zhang2018,Gong2019}.
{In a nonviscous fluid, 
the radiation torque on a spherical particle only occurs 
if the particle absorbs acoustic energy~\cite{Silva2012}.}
Albeit, nonabsorbing particles without spherical symmetry may develop the radiation torque.
Notable examples are microfibers~\cite{Schwarz2015} and nanorods~\cite{Wang2012}.
Some numerical methods have been employed to study 
the radiation torque on spheroids~\cite{Wijaya2015,Jerome2019}.

Despite the importance of the aforementioned numerical studies,
they do not reveal the full physical picture of the acoustic radiation torque.
Also, no investigation on the acoustic spin transfer to a 
spheroidal particle in a viscous fluid was performed to date.
We are not the first to theoretically investigate the acoustic radiation torque effects on spheroids. 
However, the previous work by Fan \textit{et al.}~\cite{Fan2008} 
is mainly devoted to developing a general theoretical scheme 
for arbitrarily shaped particles.

The goal of this paper is to put the acoustic radiation torque on a  spheroidal particle in a new perspective by establishing its connection with the acoustic spin.
To this end, we consider a first-order Bessel vortex beam (FOBB) in broadside incidence to a subwavelength spheroidal particle in the beam axis.
{Our choice relies on the fact that the acoustic FOBB possesses spin, which corresponds to the local expectation value of a spin-1 operator~\cite{Bliokh2019}. 
This beam not only may produce a radiation torque on the particle but also a time-averaged force, known as the acoustic force~\cite{Silva2013,Zhang2011,Leao-Neto2016}. 
Besides, some symmetry considerations have motivated the choice for a prolate spheroidal particle. This object has axial symmetry (i.e., it is invariant to a rotation around the major axis). In particle physics terms, we may classify the prolate spheroid as a spin-0 particle concerning axial rotations. 
On the other hand, rotations around the minor axis (transverse rotations) can be described by the interfocal vector, which has a $2\pi$ rotational symmetry. 
Under this circumstance, the prolate spheroid can be regarded as a spin-1 particle. 
At this point, we contemplate that the FOBB spin can only 
induce a transverse spin on the spheroid, which is a spin-1 particle.
}

{Our paper is outlined as follows.} First, we calculate the spin of a Bessel beam.
Afterward, we obtain the radiation torque considering a nonviscous fluid by solving the related scattering problem in spheroidal coordinates and integrating the result in a far-field spherical surface. 
We then establish the spin-torque relation and obtain simple asymptotic expressions of the torque as the particle geometry approaches a sphere and an infinitely thin spheroid.
Assuming a Stokes' flow as the particle spins around its minor axis~\cite{Kong2012}, we derive the relation between the acoustic spin and angular velocity.
We predict the angular velocity of microparticles
using the typical parameter values of the acoustic levitation~\cite{Marzo2015} and
acoustical vortex tweezer~\cite{Baudoin2019} devices.
Additionally, the theoretical predictions are in excellent agreement with finite-element results of the radiation torque.

\section{Acoustic spin} 
\label{sec:spin}
\begin{figure}
	\centering
	\includegraphics[scale=.27]{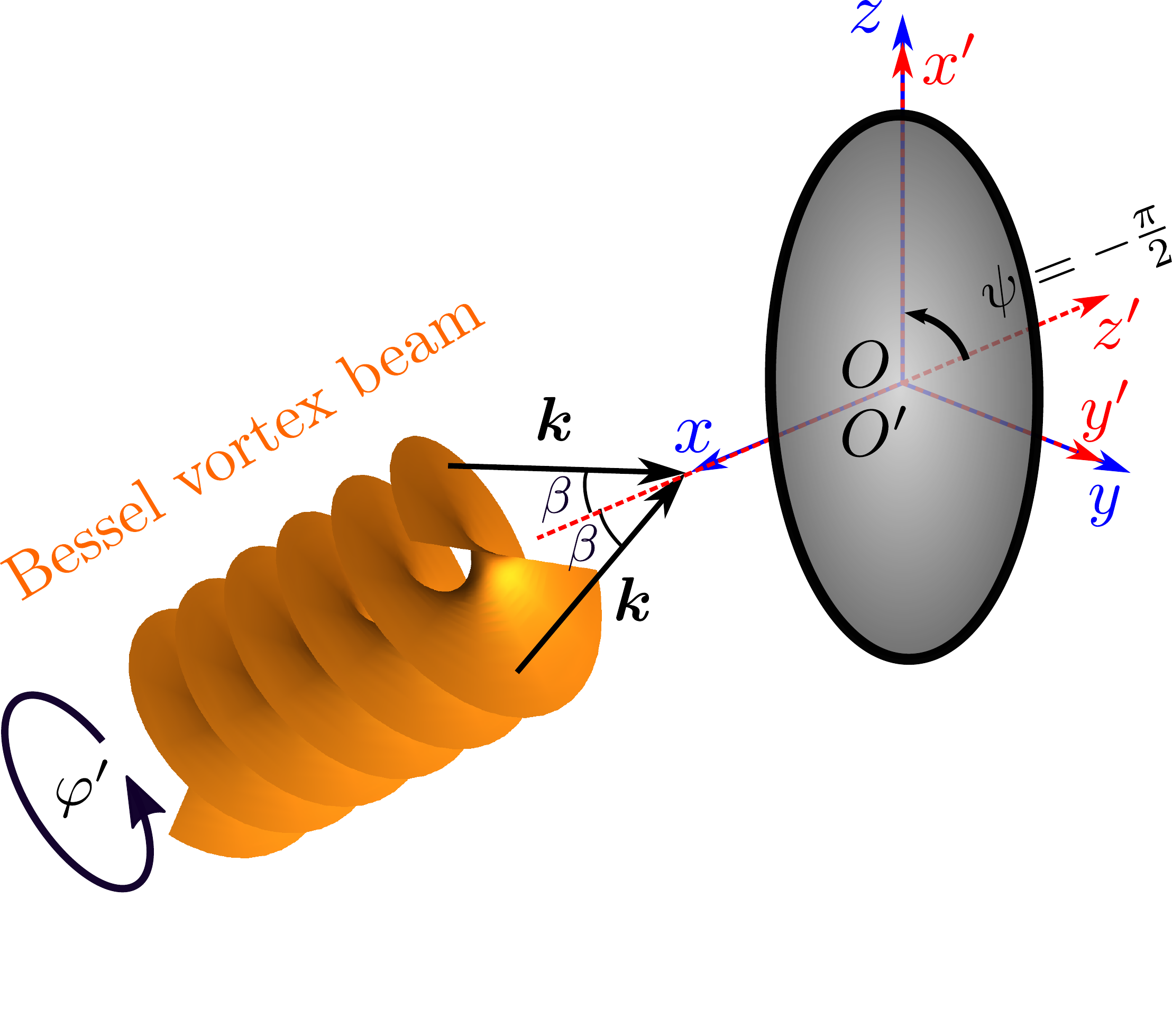}
	\caption{ 
		\label{fig:BB_sketch}
		A Bessel vortex beam of $\ell$th-order 
		with  half-cone angle $\beta$ interacting with a spheroidal particle.
		The beam propagates along the $x$ axis toward $-\infty$
		and along the $z'$ axis toward $+\infty$.
		The center of both coordinate systems $O(x,y,z)$ (blue axes)  and $O'(x',y',z')$ (red axes) are located in the particle geometric center.
		A $90^\circ$-counterclockwise rotation around $y$ and $y'$ axis maps $O'$ onto $O$ system. 
	}
\end{figure}

Assume that a Bessel vortex beam of order $\ell$ (also known as vortex charge) and angular frequency $\omega$
propagates in fluid of density $\rho_0$, adiabatic speed of sound $c_0$, and compressibility $\beta_0=1/\rho_0 c_0^2$.
The beam interacts with a subwavelength prolate spheroidal particle,
e.g. the particle dimensions are much smaller
than the acoustic wavelength.
A fixed laboratory coordinate system $O'$ coincide to the particle center which lies in the beam axis as depicted in Fig.~\ref{fig:BB_sketch}.

In the laboratory system,
the incident Bessel beam is described in
cylindrical coordinates ($\varrho', \varphi', z'$) by the velocity potential 
\begin{equation}
\label{Besselbeam}
\phi_{\textrm{in}}=\phi_0 J_{\ell}(k\varrho'\sin \beta) \textrm{e}^{\ii kz'\cos\beta} \textrm{e}^{\ii \ell \varphi'},
\end{equation}
where `i' is the imaginary unit, $\phi_0 =p_0/ k \rho_0 c_0$
(with $p_0$ being the beam peak pressure) is the potential magnitude, 
$J_\ell$ is the cylindrical Bessel function of $\ell$th-order,
$k=\omega/c_0$,  
$\beta$ is the beam half-cone angle.
The beam wavevector is $\vec{k} = k \left(\sin\beta \, \vec{e}_{\varrho'} +  \cos\beta\, \vec{e}_{z'}\right)$,
with $\vec{e}_{\varrho'}$ and $\vec{e}_{z'}$ being the radial and axial unit vectors.
The time-dependent term $\ee^{-\ii\omega t}$ is omitted for simplicity.
The incident pressure and velocity fields are given, respectively, by $\vec{v}_\text{in}=\nabla \phi_\text{in}$
and
$p_\text{in} = \ii k \rho_0 c_0 \phi_\text{in}$.

The acoustic  spin  density  of the incident beam is defined by~\cite{Bliokh2019}
\begin{equation}
\label{spin}
\vec{S} = \frac{\rho_0}{2 \omega}\text{Im}\left[\vec{v}_\text{in}^*\times
\vec{v}_\text{in}\right],
\end{equation}
where `Im' means the imaginary-part of a quantity.
The acoustic spin is an intrinsic local property of the  beam.
Inasmuch as the velocity field is irrotational $\nabla\times
\vec{v}_\text{in} = \vec{0}$, the spin satisfies the conservation law $\nabla \cdot \vec{S} = 0$.

By substituting Eq.~\eqref{Besselbeam} into Eq.~\eqref{spin}, we find
the axial spin as
\begin{equation}
\label{Szp}
{S}_{z'}(\varrho') = \frac{2 E_0 \sin \beta}{ \omega k r} 
J_{\ell}(k \varrho' \sin\beta )
\dot{J}_{\ell}(k \varrho' \sin\beta ),
\end{equation}
where 
$E_0 = \beta_0 p_0^2/2$ is the characteristic energy density of the beam, and the dot means derivative with respect
to the function's argument.


The only case where the on-axis spin is not zero corresponds
to a FOBB $(|\ell|= 1)$.
For simplicity we consider $\ell=1$.
Referring to Eq.~\eqref{Szp}, the axial spin  is given by
\begin{equation}
\label{Szprime}
		\vec{S}_{z'}(0) =  \frac{E_0\sin^2\beta}{2\omega}\,\vec{e}_{z'},
\end{equation}		
We note that the beam energy $E_0$ can be assessed by
measuring the acoustic spin of the FOBB.

\section{Scattering in the long-wavelength limit}
The spheroidal particle  has a major and minor axis denoted by $2a$ and $2b$, respectively, with  interfocal distance being 
$d=2 \sqrt{a^2-b^2}$.
The acoustic scattering is now described
in a coordinate system $O$ fixed in the geometric center  of the particle at rest.
The major axis lies in the $z$ direction--see Fig.~\ref{fig:BB_sketch}.
For symmetry reasons, we describe the particle in prolate spheroidal coordinates to which 
$\xi\geq 1$ is radial distance,
$-1\leq\eta\leq1$, and  $0\leq\varphi\leq2\pi$ is azimuth angle. 
In this case, 
the particle corresponds to the surface defined by
$\xi = \xi_0 = 2a/d=\text{const.}$
The particle aspect ratio is defined as the major-to-minor axis ratio, which relates to the particle geometric parameter $\xi_0$ as
\begin{equation}
\frac{a}{b} = \frac{1}{\sqrt{1-\xi_0^{-2}}}.
\end{equation}
The particle volume is 
$
V=4\pi ab^2/3=
\pi d^3 \xi_0 (\xi_0^2 -1)/6.
$
A spherical particle
of radius $r_0$ 
 is recovered as
 $\xi_0\rightarrow \infty$, with
 $\xi_0 d\rightarrow 2 r_0$.
Whereas
a slender  particle 
corresponds to the minor semiaxis being much smaller than the major semiaxis, $a/b\gg 1$ and then
$\xi_0\approx 1$.

In the long-wavelength  scattering analysis,
we define the expansion parameter as
proportional to the interfocal-to-wavelength ratio as
\begin{equation}
\label{kd}
\epsilon=\frac{kd}{2}=\frac{ka}{\xi_0}\ll 1.
\end{equation}
{We emphasize that the other size parameter related to 
the minor semiaxis $b$, say $kb/\xi_0$, is also much smaller than one, as $b<a$.}
In this case, only the monopole and dipole modes of the incident and scattered waves are
needed to describe the particle-wave interaction.
Accordingly, the partial wave expansions of the incident and scattering potential velocities {in the particle frame} are given in prolate spheroidal coordinates by~\cite{Flammer2005}
\begin{subequations}
	\label{phi}
	\begin{align}
	\phi_\text{in} &= \phi_0 \sum_{n=0}^1 \sum_{m=-n}^n  a_{nm}S_{nm}(\epsilon,\eta) R_{nm}^{(1)}(\epsilon,\xi)\ee^{\ii m\varphi},
	\label{pin}\\
	\phi_\text{sc} &= \phi_0 \sum_{n=0}^1 \sum_{m=-n}^n  a_{nm}s_{nm}S_{nm}(\epsilon,\eta) R_{nm}^{(3)}(\epsilon,\xi)\ee^{\ii m\varphi},
	\label{psc}
	\end{align}
\end{subequations}
where  $S_{nm}$ is the angular function of the first kind, and $R_{nm}^{(1)}$ and $R_{nm}^{(3)}$  are the radial functions of the first and third kind, respectively.
The quantities $a_n^m$ and $s_{nm}$ are the beam-shape and  scaled scattering coefficients.

We assume that the particle behaves as a rigid and immovable spheroid.
Hence, the velocity normal component is zero on the particle surface, $\partial_{\xi}(\phi_\inc+\phi_\sca)_{\xi=\xi_0}=0$.
Using \eqref{phi} in this condition, one obtains
the scattering coefficient as
\begin{equation}
s_{nm}=- \frac{\partial_{\xi} R_{nm}^{(1)}}{\partial_{\xi}R_{nm}^{(3)}}
\biggr|_{\xi=\xi_0}.
\label{coef:scattering}
\end{equation}

We shall see in Sec.~\ref{sec:ART} that in the long-wavelength limit, only the dipole scattering coefficients
contribute to the acoustic radiation torque.
Hence, after Taylor-expanding 
the radial functions given in \eqref{Rnms}
around $\epsilon=0$, we obtain 
the dipole scattering coefficients as~\cite{Silva2018}
\begin{subequations}
	\label{scatt_spheroid}
	\begin{align}
	s_{10} &=\frac{\ii\epsilon^3}{6}f_{10}-\frac{\epsilon^6}{36}f_{10}^2 ,\\
	s_{1,-1}
	&=s_{11}=\frac{\ii\epsilon^3}{12}f_{11}-\frac{\epsilon^6}{144}f_{11}^2,
	\end{align}
\end{subequations}
where 
\begin{subequations}
		\label{factors}
	\begin{align}
	f_{10} &=\frac{2}{3}\left[\frac{\xi_0}{\xi_0^2-1}-\ln\left(\frac{\xi_0+1}{\sqrt{\xi_0^2-1}}\right)\right]^{-1} ,\\
	f_{11}
	&=\frac{8}{3}\left[\frac{2-\xi_0^2}{\xi_0(\xi_0^2-1)}+\ln\left(\frac{\xi_0+1}{\sqrt{\xi_0^2-1}}\right)\right]^{-1}
	\end{align}
\end{subequations}
are the scattering factors.

In the far-field $k\xi\gg 1$, 
the spheroidal expansion in \eqref{phi} asymptotically approaches the expansion in spherical coordinates $(r,\theta, \varphi)$ as follows~\cite{Silva2018}
\begin{subequations}
	\label{far}
	\begin{align}
	\phi_\inc &= \frac{\phi_0}{kr}
	\sum_{n=0}^1\sum_{m=-n}^n {a}_{nm}
	\sin\left(kr -\frac{n\pi }{2}\right)Y_n^m(\theta,\varphi),
	\\
	\phi_\sca &=\phi_0
	\frac{\ee^{\ii kr}}{kr} 
	\sum_{n=0}^1\sum_{m=-n}^n\ii^{-n-1}{a}_{nm}s_{nm}
	Y_n^m(\theta,\varphi),
	\end{align}
\end{subequations}
where 
$
Y_n^m(\theta,\varphi)$
is the spherical harmonic of $n$th-order and $m$th-degree.
Here the beam-shape coefficient $a_{nm}$ describes an incident wave in spherical coordinates.
Hereafter, we shall consider the beam-shape coefficients in spherical coordinates.

\section{Acoustic radiation torque}
\label{sec:ART}
The density of linear momentum flux carried by an acoustic wave is well-known from the fluid  mechanics theory~\cite{Silva2011}
$\overline{\bf{P}} = -\overline{\mathcal{L}}{\bf I} + \rho_0 \overline{\vec{v}\vec{v}}$,
with the over bar denoting time-average over a wave period and
$\bf I$ being the unit tensor.
The acoustic fields $\mathcal{L}$ and $ \rho_0 \vec{v}\vec{v}$ are the 
 Lagrangian density and Reynolds' stress tensor.
The  density of angular momentum flux is then
$\overline{\bf{L}}=\vec{r}\times \overline{\bf{P}}$.
The radiation force exerted by the incident wave on an surface element $\dd S$ of the particle is
$\dd \vec{F}_\text{rad} =  \overline{\bf{P}} \cdot \vec{n} \, \dd S$, with $\vec{n}$ being the outwardly unit vector at
the particle surface $S_0$,
whereas the moment of the infinitesimal radiation force is given by
$\dd\vec{\tau}_\text{rad} = \vec{r} \times \dd \vec{F}_\text{rad}
= \overline{\bf{L}}\cdot \vec{n} \,\dd S $.
Therefore, the acoustic radiation torque on the particle is expressed by
\begin{equation}
\label{Ni}
\vec{\tau}_{\text{rad}} = \int_{S_0} \overline{\bf{L}}\cdot \vec{n} \, \dd S.
\end{equation}
As the angular momentum flux satisfies the conservation law~\cite{Zhang2011c} $\nabla \cdot \overline{\bf{L}} = \vec{0}$,
the integral can be evaluated over a virtual surface $S_1$ of a sphere in the far-field $kr\gg 1$ 
that encloses the particle.
Accordingly,  the radiation torque is expressed by
$
\vec{\tau}_{\text{rad}} = -\int_{S_1} \left(\vec{r}\times
\rho_0 
\overline{ \vec{v} \vec{v}}\right)\cdot \vec{e}_r \, \dd S
$, with $\vec{e}_r$ being the unit-vector in radial direction.
The fluid velocity is the sum of the incident and scattered velocities,
$\vec{v}=\vec{v}_\text{in} + \vec{v}_\text{sc}$.
Substituting the total fluid velocity into 
the far-field expression of the radiation torque 
and noting that $\overline{\vec{v}\vec{v}}= (1/2)\text{Re}[\vec{v}\vec{v}^*]$,
we find~\cite{Lopes2016} 
\begin{equation}
\label{Nend}
\vec{\tau}_{\text{rad}} = -\frac{\rho_0 r^2}{2}
\text{Re} \int_{\Omega_\text{s}}  
\vec{r}\times(\vec{v}_\text{in} \vec{v}_\text{sc}^*+ \vec{v}_\text{sc}\vec{v}_\text{in}^* + \vec{v}_\text{sc}\vec{v}_\text{sc}^*)\cdot \vec{e}_r\, \dd \Omega_\text{s},
\end{equation} 
where `Re' means the real part of, 
the asterisk denotes complex conjugation,  $\Omega_\text{s}$ represents the unit-sphere,
and $\dd \Omega_\text{s}$ is solid angle.
No torque is formed in the absence of the particle; hence,
$\text{Re}\int_{\Omega_\text{s}}\vec{r} \times \vec{v}_\text{in} \vec{v}_\text{in}^*\cdot \vec{e}_r\,\dd\Omega_\text{s} = \vec{0}$.
Using the partial wave expansion  in the far-field  as given in \eqref{far} into Eq.~\eqref{Nend}, one can show that
the Cartesian coordinates of the radiation torque is expressed by~\cite{Silva2012}
	\begin{subequations}
		\label{torque_rayleigh}
		\begin{align}
		\nonumber
		{\tau}_{\text{rad},x} &= - \frac{E_0 }{k^3\sqrt{2}}\,\text{Re} 
		\biggl[
		(a_{1,-1}+a_{11})(1+s_{11})a^*_{10}s^*_{10}\\
		&+
		a_{10}(1+s_{10})(a_{1,-1}^* + a_{11}^*)
		s_{11}^*
		\biggr],\\
		\nonumber
		{\tau}_{\text{rad},y} &=  - \frac{E_0 }{k^3\sqrt{2}}\,\text{Re} \biggl[\ii\,
		(a_{1,-1}-a_{11})(1+s_{11})a^*_{10}s^*_{10}\\
		&-\ii\,
		a_{10}(1+s_{10})(a_{1,-1}^* -a_{11}^*)
		s_{11}^*
		\biggr],\\
		{\tau}_{\text{rad},z} &= \frac{E_0 }{k^3 }\,\text{Re}
		\left[(|a_{1,-1}|^2 - |a_{11}|^2)
		(1+s_{11})s_{11}^*
		\right].
		\end{align}
\end{subequations}
Clearly, the  radiation torque is caused by the  nonlinear interaction between the incident and scattered dipole modes.

To compute the radiation torque from \eqref{torque_rayleigh},
the beam-shape coefficients of the incident wave should be known 
\textit{a priori}.
Notable examples are plane waves, and Bessel vortex~\cite{Gong2017} and  Gaussian beams~\cite{Mitri2014}.
Numerical schemes and the addition theorem of spherical functions have  been employed to compute the coefficients 
for different types of beam~\cite{Silva2011a,Mitri2011,Silva2015a,Silva2013,Lopes2016}.

We now proceed to calculate the FOBB radiation torque 
in broadside incidence to the  particle.
In this case, the Bessel beam propagates along the $x$ axis toward $-\infty$ in the particle  system $O$.
We see in Fig.~\ref{fig:BB_sketch} that the laboratory system $O'$ can be mapped onto
the particle system $O$ through a $90^\circ$-counterclockwise rotation around the $y'$ axis. 

In the laboratory system $O'$, the beam-shape coefficient of the FOBB is given by~\cite{Mitri2011}
\begin{equation}
\label{BSC}
a'_{n\ell}= 4\pi\, \ii^{n-m} Y_n^{m}(\beta, 0)H(n-m)\delta_{m\ell},
\end{equation}
where $H(n-m)$ is the unit-step function, which is equal to $0$ for $n-m<0$ and $1$ for $n-m\ge0$.
According to~\eqref{torque_rayleigh},
we have to compute the dipole beam-shape coefficient $a_{1,m}$ in the particle system $O$.
%
The relation between the beam-shape coefficient in the laboratory and particle system is given through the Wigner $D$-function $D_{m\mu }^{n}(\alpha,\psi,\zeta)$ as~\cite{Mishchenko2002}
\begin{equation}
\label{BSC1}
a_{nm}= \sum_{\mu=-n}^n a'_{n\mu} D_{m\mu }^n\left(\alpha,\psi,\zeta\right),
\end{equation}
where $\alpha$, $\psi,$ and $\zeta$ are the Euler angles.
Mapping system $O$ onto $O'$ corresponds to the Euler angles $\alpha=0$, $\psi=-\pi/2$, and $\zeta=0$.
To obtain $a_{1,m}$ we need only the dipole beam-shape coefficient in system $O'$, $a'_{1,\mu}$. 
According to Eq.~\eqref{BSC1}, this coefficient is
$a_{1,\mu}' = -\delta_{\mu,1}\sqrt{6 \pi} \sin\beta.$
By replacing it into Eq.~\eqref{BSC1}, we find the dipole beam-shape coefficients in
the particle system as
\begin{equation}
\label{BSBB}
a_{1,-1} = a_{11} = -\sqrt{\frac{3 \pi}{2}}\sin \beta,\quad
a_{10} = \sqrt{{3 \pi} }\sin \beta.
\end{equation}
Using this result into \eqref{torque_rayleigh},
we find the radiation torque along the $x$-axis as
\begin{equation}
{\tau}_{\text{rad},x} = \frac{3\pi  }{k^3}E_0\sin^2\beta\,\text{Re} 
\left[s_{10}+ s_{11}+ 2 s_{10}s_{11}^* 
\right].
\end{equation}
Using the scattering coefficients of \eqref{scatt_spheroid}
into this expression, we find 
\begin{subequations}
	\begin{align}
\label{tau_bvb}
\vec{\tau}_{\text{rad}} &=  -   (ka)^3  \chi \pi a^3     E_0\sin^2\beta\,\vec{e}_x,\quad ka\ll1,\\
	\chi &=  \frac{ \left(f_{11} - 2f_{10}\right)^2}{48\xi_0^6},
\end{align}
\end{subequations}
where $\chi$ is {related to the difference of the dipole factors}.

The asymptotic gyroacoustic expressions
as the particle geometric parameter describes a spherical ($\xi_0\gg1$) and slender 
particle ($\xi_0\approx 1$) are given,  respectively, by
\begin{subequations}
	\label{asymp_gamma}
	\begin{align}
	\label{gamma_sphere}
	\chi &=
	\frac{3}{400}\left(
	\frac{1}{ \xi_0^{4}}
		- \frac{47}{35\xi_0^{6}}\right),\\
	\chi & =
	\frac{4}{27}(\xi_0 -1)^2 + {\frac{4}{9}}(\xi_0 -1)^3 
	\left(3 + \ln\left[\frac{(\xi_0-1)^2}{4}\right] \right).
	 \label{gamma_slender}
	\end{align}
\end{subequations}
The radiation torque vanishes as the
particle geometry approaches a sphere, $\lim_{\xi_0\rightarrow \infty} \tau_\text{rad} = 0$.
This is supported by the fact that no torque is produced on a nonabsorbing sphere~\cite{Silva2012}.

Importantly, both asymptotic expansions of the gyroacoustic factor in \eqref{asymp_gamma} approach to zero.
This suggests that the geometric torque factor $\chi$ should have an extreme value in the interval $1\le\xi_0<\infty$.
Using the Nelder-Mead numerical method {through  \texttt{NMaximize} function of Mathematica Software~\cite{Mathematica}}, we find the maximum value $\chi_\text{max} = 0.14$ at $\xi_0=1.3181$.

We now establish the connection between the acoustic radiation torque and acoustic spin.
To do so, we  express the radiation torque in the
laboratory frame (system $O'$) as
$
\vec{\tau}_{\text{rad}} =    \gamma \pi a^3    (ka)^3  E_\text{in}(0)\,\vec{e}_{z'}.
$
The torque is positive given that $\vec{e}_{z'} = -\vec{e}_x$, i.e., the $z'$ and $x$ axis have
opposite orientation.
Using Eq.~\eqref{Szprime}, we find at the spin-induced radiation torque on the particle as
\begin{equation}
\label{tauspin}
	{\tau}_\text{rad} = \frac{ \chi\pi a^3\omega}{2 }
	{S}_{z'}(0).
\end{equation}

%

\section{Particle angular velocity}
In broadside incidence, a FOBB may set the  spheroidal particle to spin around its minor axis.
Here  we consider that the particle is immersed in a viscous incompressible fluid with dynamic viscosity $\mu_0$.
To simplify our analysis, we assume that the yielded flow due to the particle spin  has a small  Reynolds number $Re \ll 1$,
i.e., the so-called Stokes' flow. 
It is worth noticing that by solving the acoustic scattering problem we have considered a compressible fluid. 

In the laboratory frame (system $O'$), the rotation dynamics is described by the Newton's second law,
\begin{equation}
\label{rot_dynamics}
I_\text{p} \dot{\Omega} =   \tau_\text{rad} - \tau_\text{drag}(\Omega),
\end{equation}
where $I_\text{p}$ is the particle moment of inertia relative to the minor  axis,
$\Omega$ is the particle angular velocity,
and $\tau_\text{drag}$ is the drag torque that counteracts the radiation torque.

Assuming the no-slip boundary condition at the particle surface $\xi=\xi_0$, one can find the drag torque as~\cite{Kong2012}
\begin{subequations}
	\label{Ndragall}
\begin{align}
\label{Ndrag}
&\tau_\text{drag} = \pi \mu_0 d^3  \tilde{\tau}_\text{drag}\, {\Omega} ,\\
&\tilde{\tau}_\text{drag} =
{\frac{4}{3}\frac{1- 2\xi_0^2}{2\xi_0- (1+\xi_0^2)\ln\left(\frac{\xi_0+1}{\xi_0-1}\right)}.}
\label{taudrag}
\end{align}
\end{subequations}
The dimensionless drag torque $\tilde{\tau}_\text{drag}$ depends only of the geometry of the particle.
As the particle  asymptotically approaches ($\xi_0\gg 1$) a sphere of radius $r_0$,
we recover the classical result of the drag torque for a spherical geometry,
$
{\tau}_\text{drag}^\text{sphere} = 8 \pi \mu_0 r_0^3 \Omega,
$
with $r_0\approx \xi_0 d/2$.
\begin{table}
	\caption{The acoustic parameters of acoustofluidics and levitation systems at room temperature~\cite{Lide2004}.
		\label{tab:param}
	}
	\begin{tabular}{lccc}
		\hline\hline
		Medium &
		\begin{tabular}[c]{@{}c@{}}Density\\  $\rho_0\,\si{[\kilogram\per\meter\cubed}]$\end{tabular}
		&
		\begin{tabular}[c]{@{}c@{}}Speed of sound \\  $c_0\,\si{[\meter\per\second}]$\end{tabular}
		&
		\begin{tabular}[c]{@{}c@{}}Dynamic viscosity \\  $\mu_0\,[\si{\pascal\cdot\second}]$\end{tabular}
		\\
		\hline
		Air    & $1.22$& $343$ & $1.86\times10^{-5}$ \\
		Water & $998$ & $1483$& $10^{-3}$	\\
		\hline
		\hline
	\end{tabular}
\end{table}
\begin{figure*}
	\centering
	\includegraphics[scale=.48]{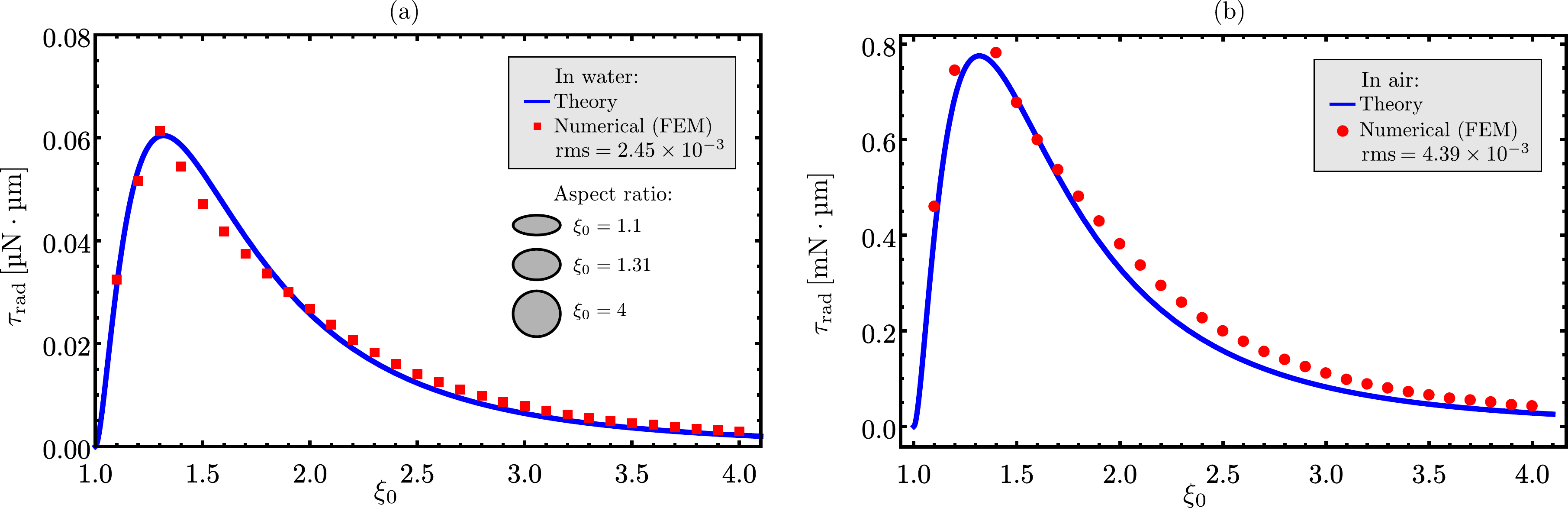}
	\caption{ 
		The radiation torque exerted on a particle in (a) water and (b) air as a function of the particle geometric parameter $\xi_0$.
		The torque is evaluated  with Eq.~\eqref{tau_bvb} with
		$\beta=\pi/4$.
		The parameters for water are  $a=120\,\mu$m, $f=1\,$MHz, and $p_0=500\,$kPa; while for air, we have
		$a=680\,\mu$m, $f=40\,$kHz, and $p_0=3.5\,$kPa.
		The red square and circular dots correspond to finite-element simulation results.
		The maximum value of the radiation torque is at $\xi_0=1.31$ ($a/b=1.54$).
		Three particles with different aspect ratios ($\xi_0=1.1,1.31,4$) are depicted in panel (a).
		\label{fig:torque}
	}
\end{figure*}

As the angular velocity increases, the radiation and drag torques balance each other.
Thus, the particle reaches a stationary angular velocity
that can be obtained by combining
 Eqs.~\eqref{tau_bvb}, \eqref{rot_dynamics} and \eqref{Ndragall}, 
\begin{subequations}
\begin{align}
\label{omegap}
\vec{\Omega}_\text{st}&=  \frac{\vec{\tau}_\text{rad}}{\pi\mu_0 d^3  \tilde{\tau}_\text{drag}}=
(k a)^3 \tilde{\Omega}_\text{st}
\frac{ E_0}{ \mu_0 }\sin^2\beta
\,\vec{e}_{z'},\\
\tilde{\Omega}_\text{st} &= \frac{\chi \xi_0^3}{{8}\tilde{\tau}_\text{drag}},
\label{omegatil}
\end{align}
\end{subequations}
with $\tilde{\Omega}_\text{st}$ being the dimensionless angular velocity.
By measuring the angular velocity $\Omega_\text{st}$ and knowing the particle and beam parameters, one can determine the fluid viscosity $\mu_0$ through Eq.~\eqref{omegap}.

For a quasispherical and slender particle, the dimensionless angular velocity is, respectively,
\begin{subequations}
\begin{align}
\tilde{\Omega}_\text{st} &=
\frac{3}{3200}\left(
 \frac{1}{\xi_0^{4}} - \frac{31}{70 \xi_0^{6}}\right),
 \quad \xi_0\gg 1,\\
\tilde{\Omega}_\text{st} &=-
\frac{1}{36}(\xi_0-1)^2\left[1 + \ln\left(\frac{\xi_0-1}{2}\right)
\right].
 \quad \xi_0\approx 1.
\end{align}
\end{subequations}
%



%

The relation between the axial acoustic spin and particle angular velocity follows by replacing  Eq.~\eqref{tauspin} into \eqref{omegap},
\begin{subequations}
\begin{align}
\label{SxO}
\vec{S}_{z'}(0) &=   \gamma \vec{\Omega}_\text{st} \\
\gamma &= 
\frac{16 \mu_0 \tilde{\tau}_\text{drag}}{ (k a)^3 \chi \xi_0^3 \omega},
\label{chi_OS}
\end{align}
\end{subequations}
where $\gamma$ is the \textit{gyroacoustic} ratio
of the spin and angular velocity
in the SI units of $\si{\kg\per\meter}$.
This result describes how the spin is transferred to a subwavelength spheroidal particle.
It also enables the experimental assessment of the acoustic spin by measuring the angular velocity
of a subwavelength spheroidal particle.

\section{Model predictions}
We provide some  predictions for typical experimental setups of
acoustical vortex tweezers~\cite{Baudoin2019} and acoustic levitation~\cite{Marzo2015}
to which the particle is immersed in a water-like medium and air, respectively.
The acoustic parameters of these fluids are summarized in Table~\ref{tab:param}. 
The particle has a fixed major semiaxis of $a=\SI{680}{\micro\meter}$ in air and $a=\SI{120}{\micro\meter}$ in water.

The theoretical predictions will be compared with 3D finite-element simulation results performed in Comsol Multiphysics (Comsol Inc., USA). 
The radiation torque was computed by numerical integration of the angular momentum flux $\overline{\bf{L}}$ over the particle surface as described in Eq.~\eqref{Ni}.  
The mean discretization length on the surface is $b/50$; while in the surrounding fluid, we consider at least $\lambda/12$. 
The domain has a cylindrical geometry with $36 b$ diameter and height.
We have also adopted the 
first-order scattering boundary condition at the domain edges.
%

In Fig.~\ref{fig:torque}, we show the radiation torque exerted on a spheroidal particle as a function of 
the geometric parameter $\xi_0$ in water and air.
The torque is evaluated  with Eq.~\eqref{tau_bvb}.
The pressure peaks are
$p_0 = \SI{3.5}{\kilo\pascal}$ (air) and $p_0 = \SI{0.5}{\mega\pascal}$ (water).
The driving frequencies are
$f=\SI{40}{\kilo\hertz}$ (air) and $f=\SI{1}{\mega\hertz}$ (water).
The half-cone angle of the beam is $\beta=\pi/4$.
According to Eq.~\eqref{kd} the size parameter $\epsilon$ is always smaller than $0.51$.
The radiation torques peak at $\xi_0=1.31$, which corresponds to the aspect ratio $a/b=1.54$.
finite-element results are also depicted for comparison.
The root mean square error (rms) is about $10^{-3}$ in both cases.

\begin{figure}
	\centering
	\includegraphics[scale=.46]{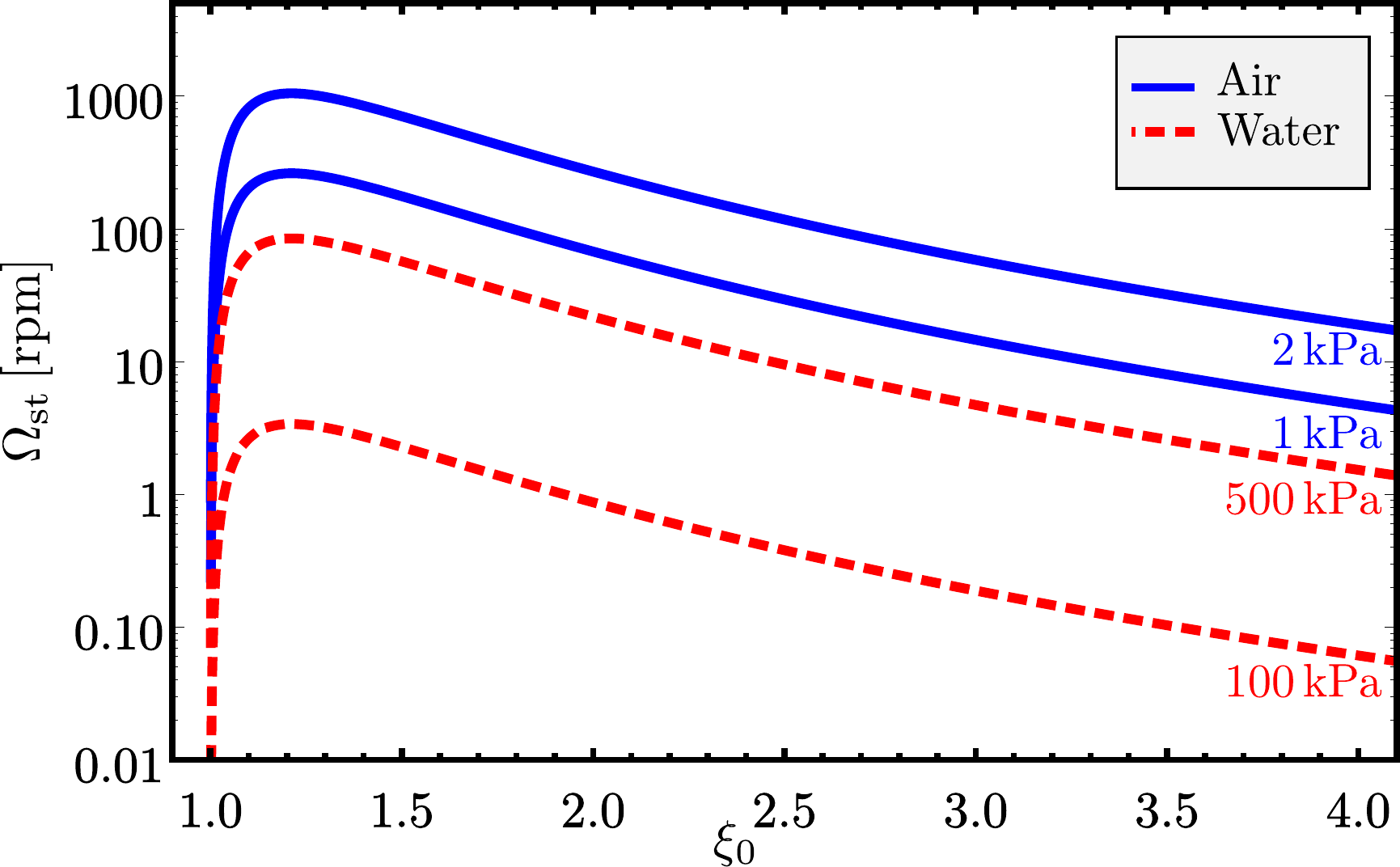}
	\caption{ 
		The stationary angular velocity as a function of the particle geometric parameter $\xi_0$ for different peak pressures in water and air.
		The velocity is calculated with Eq.~\eqref{omegap}  and $\beta=\pi/4$.
		The parameters for water are  $a=120\,\mu$m and $f=1\,$MHz; and for air,
		$a=680\,\mu$m and $f=40\,$kHz. 
		The maximum value of the angular velocity is at $\xi_0=1.21$ ($a/b=1.77$).
		\label{fig:angVel}
	}
\end{figure}

In Fig.~\ref{fig:angVel}, 
we plot the  angular velocity versus the particle geometric parameter $\xi_0$ with different peak pressures
$p_0=1,\,\SI{2}{\kilo\pascal}$ (air) and
$p_0=100,\,\SI{500}{\kilo\pascal}$ (water).
We note that the peak velocity is reached at $\xi_0=1.21$,
which corresponds to the aspect ratio $a/b=1.77$.
{
	When compared to the radiation torque, this maximum value appears for a different geometric parameter. 
	This happens because the viscous drag torque acts on the particle, as shown in Eq.~\eqref{omegatil}, changing the optimal aspect ratio for the angular velocity.}
In water, the angular velocity can be as large as $100$\,rpm, whereas in air, it can be ten times this value.

\section{Summary and conclusion}
We have demonstrated that the acoustic spin can be transferred
to a subwavelength spheroidal particle.
Using the partial wave expansion of the incident and scattered velocity potentials in spheroidal coordinates
and integrating the total angular momentum density in the far-field, 
we derived a general expression of the radiation torque in the long-wavelength limit.
Considering a broadside incidence of a FOBB onto the particle centered at the beam axis,
we obtained the corresponding radiation torque.
In turn, the torque produces an angular velocity on the particle that rotates around its minor axis.

{We offer a more fundamental explanation of the spin-induced torque using a description from quantum physics. The acoustic FOBB is regarded as a spin-1 field~\cite{Bliokh2019}, whereas a prolate spheroid can be classified as a spin-0 and spin-1 particle under axial and transverse rotations, respectively. 
Therefore, we found that the spin can only be transferred from the FOBB in broadside incidence to the particle inducing a transverse rotation.}
{Importantly,  axial rotations can be generated by viscous torques caused by tangential stresses within the particle boundary layer.
	However, the viscous torque can be neglected
	as the boundary layer thickness, $\delta = (2\mu_0/\rho_0\omega)^{1/2}$ is much smaller than the particle size~\cite{Lee1989}.
	Here $\delta/a\sim 10^{-3}$ (in water) and
	$\delta/a\sim 10^{-2}$ (in air).
	For this reason, this torque was discarded in our analysis.}

The stationary angular velocity is obtained by taking the radiation and drag torque balance in 
Eq.~\eqref{rot_dynamics}.
Considering the physical parameters of acoustofluidic and experimental levitation setups, our model predicts that
the stationary angular velocity can reach $\SI{100}{\,rpm}$ in water and $\SI{1000}{\,rpm}$ in air.
Therefore, it is feasible to measure the angular velocity and use the result to obtain the acoustic spin.
We can also determine the fluid viscosity by measuring the angular velocity.
Additionally, by measuring the acoustic spin, we can obtain
the beam energy density as described in Eq.~\eqref{spin}.
This may provide a means of assessing the energy   of focused ultrasonic vortices in acoustic levitation systems~\cite{Marzo2015} and
acoustical vortex tweezers~\cite{Baudoin2019}.

In conclusion, we have established a connection between the acoustic spin and the angular velocity of a spheroidal particle in a viscous fluid.
The developed method can also be applied to unveil the properties of other spin-carrying acoustic beams.

\begin{acknowledgments}
We thank the National Council for Scientific and Technological
Development--CNPq, Brazil (Grant No. 401751/2016-3 and
No. 307221/2016-4) for financial support.
\end{acknowledgments}


\newpage

\onecolumngrid
\appendix 
\section{Monopole and dipole radial functions}
\label{app:radfunctions}
In the long-wavelength limit,
the radial spheroidal functions  are given
to  the $\epsilon^6$-order 
by~\cite{Burke1966}
\begin{subequations}
	\label{Rnms}
\begin{align}
R_{00}^{(1)} &= 1 +
\frac{\epsilon^2}{18} \left(2-3 C_1^2\right) +\frac{\epsilon^4}{16200} \biggl[112-180 C_1^2 + 135 C_1^4
+\frac{\epsilon^2}{882} \left(2192 -8064 C_1^2 
+5670 C_1^4
-2835 C_1^6\right) \biggr],\\
R_{10}^{(1)} &=
\frac{\epsilon}{C_1}
+\frac{\epsilon^2C_1}{150}
\biggl[
2- 5C_1^2  + \frac{\epsilon^2}{4900}
\left(368-700C_1^2 + 875 C_1^4\right)
\biggr],\\
R_{11}^{(1)} &=
\frac{\epsilon S_1}{3} +
\frac{\epsilon^3S_1}{150}
\biggl[4 - 5 C_1^2 + \frac{\epsilon^2}{4900}(712-1400C_1^2 + 875 C_1^4)\biggr],
\\
\label{app:Rs}
R_{00}^{(2)} &=
-\frac{2}{\epsilon}
\biggl\{
L - \frac{\epsilon^2}{6}[6C_1+L(3C_2-5)]
+\frac{3}{5}\left(\frac{\epsilon}{6}\right)^4
\left[11 C_1 + 9 C_3+ \frac{L}{60}(1109 -1380 C_2 +135C_4)
\right]
\biggr\},\\
\nonumber
R_{10}^{(2)}&=
\frac{3 C_1}{\epsilon^2}
\biggl\{
2C_1 - \frac{C_2}{C_1} - 2L
-\left(\frac{\epsilon}{10}\right)^2 
\left[18C_1 - \frac{4C_2}{C_1} + L(22 - 10C_2)\right]
+ \frac{1}{882}\left(\frac{\epsilon}{10}\right)^4\\
&\times
\biggl[272313C1 - 864\frac{C_2}{C_1} + 7875 C_3 -
L(116073 - 99540 C_2 + 7875 C_2)\biggr]
\biggr\},
\\
\nonumber
R_{11}^{(2)} &=
-\frac{3S_1}{2 \epsilon^2}\biggl\{
\frac{C_1}{S_1^2} - 2L - \left(\frac{\epsilon}{10}\right)^4
\biggl[8C_1 \left(5-\frac{1}{S_1^2}\right) - 8L(33+5C_2)\biggr] 
-\frac{1}{196}\left(\frac{\epsilon}{10}\right)^2
\biggl[85800C_1 - 1750 C_3 
\\
&+  \frac{712 C_1}{S_1^2}
-L (106324 - 76950 C_2 - 1750 C_4)
\biggr]
\biggr\},\\
R_{nm}^{(3)}&= R_{nm}^{(1)} + \ii R_{nm}^{(2)},
\end{align}
\end{subequations}
where $R_{nm}^{(2)}$ is the radial function of the second-kind.
Note that $R_{nm}^{(i)} = R^{(i)}_{n,-m}$, with $i=1,2,3$.
The auxiliary functions are expressed by
\begin{align}
\nonumber
C_n &= \frac{1}{2}\left[
(\sqrt{\xi^2 -1} + \xi)^n
+(\sqrt{\xi^2 -1} + \xi)^{-n}
\right],\\
\nonumber
S_n &= \frac{1}{2}\left[
(\sqrt{\xi^2 -1} + \xi)^n
-(\sqrt{\xi^2 -1} + \xi)^{-n}
\right],\\
\nonumber
L &= \frac{1}{2} \ln
\left(\frac{1 + (\sqrt{\xi^2 -1} + \xi)^{-1}}
{1 - (\sqrt{\xi^2 -1} + \xi)^{-1}}
\right).
\end{align}

\section{Wigner $D$ function}
The Wigner $D$ function $D_{\mu m}^{n}(0,-\pi/2,0)$ was evaluated with Mathematica 
Software (Wolfram Inc., USA).
To the dipole approximation, we have
\begin{subequations}
	\label{Dfunction}
	\begin{align}
	&D_{00}^0 = 1, \\
	&D_{-1,-1}^{1} =D_{-1,1}^{1} =D_{1,-1}^{1} =
	D_{11}^{1} =
	 \frac{1}{2},\\
	&D_{-1,0}^{1}= D_{01}^{1} = 
	-\frac{1}{\sqrt{2}},\\
	&D_{0,-1}^{1} =D_{10}^{1}=\frac{1}{\sqrt{2}}, \\
&	D_{00}^{1} = 0. 
	\end{align}
\end{subequations}

\twocolumngrid

\begin{thebibliography}{38}%
	\makeatletter
	\providecommand \@ifxundefined [1]{%
		\@ifx{#1\undefined}
	}%
	\providecommand \@ifnum [1]{%
		\ifnum #1\expandafter \@firstoftwo
		\else \expandafter \@secondoftwo
		\fi
	}%
	\providecommand \@ifx [1]{%
		\ifx #1\expandafter \@firstoftwo
		\else \expandafter \@secondoftwo
		\fi
	}%
	\providecommand \natexlab [1]{#1}%
	\providecommand \enquote  [1]{``#1''}%
	\providecommand \bibnamefont  [1]{#1}%
	\providecommand \bibfnamefont [1]{#1}%
	\providecommand \citenamefont [1]{#1}%
	\providecommand \href@noop [0]{\@secondoftwo}%
	\providecommand \href [0]{\begingroup \@sanitize@url \@href}%
	\providecommand \@href[1]{\@@startlink{#1}\@@href}%
	\providecommand \@@href[1]{\endgroup#1\@@endlink}%
	\providecommand \@sanitize@url [0]{\catcode `\\12\catcode `\$12\catcode
		`\&12\catcode `\#12\catcode `\^12\catcode `\_12\catcode `\%12\relax}%
	\providecommand \@@startlink[1]{}%
	\providecommand \@@endlink[0]{}%
	\providecommand \url  [0]{\begingroup\@sanitize@url \@url }%
	\providecommand \@url [1]{\endgroup\@href {#1}{\urlprefix }}%
	\providecommand \urlprefix  [0]{URL }%
	\providecommand \Eprint [0]{\href }%
	\providecommand \doibase [0]{http://dx.doi.org/}%
	\providecommand \selectlanguage [0]{\@gobble}%
	\providecommand \bibinfo  [0]{\@secondoftwo}%
	\providecommand \bibfield  [0]{\@secondoftwo}%
	\providecommand \translation [1]{[#1]}%
	\providecommand \BibitemOpen [0]{}%
	\providecommand \bibitemStop [0]{}%
	\providecommand \bibitemNoStop [0]{.\EOS\space}%
	\providecommand \EOS [0]{\spacefactor3000\relax}%
	\providecommand \BibitemShut  [1]{\csname bibitem#1\endcsname}%
	\let\auto@bib@innerbib\@empty
	\bibitem [{\citenamefont {Ohanian}(1986)}]{Ohanian1986}%
	\BibitemOpen
	\bibfield  {author} {\bibinfo {author} {\bibfnamefont {H.~C.}\ \bibnamefont
			{Ohanian}},\ }\href@noop {} {\bibfield  {journal} {\bibinfo  {journal} {Am.
				J. Phys.}\ }\textbf {\bibinfo {volume} {54}},\ \bibinfo {pages} {500}
		(\bibinfo {year} {1986})}\BibitemShut {NoStop}%
	\bibitem [{\citenamefont {Belinfante}(1939)}]{Belinfante1939}%
	\BibitemOpen
	\bibfield  {author} {\bibinfo {author} {\bibfnamefont {F.~J.}\ \bibnamefont
			{Belinfante}},\ }\href@noop {} {\bibfield  {journal} {\bibinfo  {journal}
			{Physica}\ }\textbf {\bibinfo {volume} {6}},\ \bibinfo {pages} {887}
		(\bibinfo {year} {1939})}\BibitemShut {NoStop}%
	\bibitem [{\citenamefont {Shi}\ \emph {et~al.}(2019)\citenamefont {Shi},
		\citenamefont {Zhao}, \citenamefont {Long}, \citenamefont {Yang},
		\citenamefont {Wang}, \citenamefont {Chen}, \citenamefont {Ren},\ and\
		\citenamefont {Zhang}}]{Shi2019}%
	\BibitemOpen
	\bibfield  {author} {\bibinfo {author} {\bibfnamefont {C.}~\bibnamefont
			{Shi}}, \bibinfo {author} {\bibfnamefont {R.}~\bibnamefont {Zhao}}, \bibinfo
		{author} {\bibfnamefont {Y.}~\bibnamefont {Long}}, \bibinfo {author}
		{\bibfnamefont {S.}~\bibnamefont {Yang}}, \bibinfo {author} {\bibfnamefont
			{Y.}~\bibnamefont {Wang}}, \bibinfo {author} {\bibfnamefont {H.}~\bibnamefont
			{Chen}}, \bibinfo {author} {\bibfnamefont {J.}~\bibnamefont {Ren}}, \ and\
		\bibinfo {author} {\bibfnamefont {X.}~\bibnamefont {Zhang}},\ }\href@noop {}
	{\bibfield  {journal} {\bibinfo  {journal} {Nat. Sci. Rev.}\ }\textbf
		{\bibinfo {volume} {6}},\ \bibinfo {pages} {707} (\bibinfo {year}
		{2019})}\BibitemShut {NoStop}%
	\bibitem [{\citenamefont {Bliokh}\ and\ \citenamefont
		{Nori}(2019)}]{Bliokh2019}%
	\BibitemOpen
	\bibfield  {author} {\bibinfo {author} {\bibfnamefont {K.~Y.}\ \bibnamefont
			{Bliokh}}\ and\ \bibinfo {author} {\bibfnamefont {F.}~\bibnamefont {Nori}},\
	}\href@noop {} {\bibfield  {journal} {\bibinfo  {journal} {Phys. Rev. B}\
		}\textbf {\bibinfo {volume} {99}},\ \bibinfo {pages} {174310} (\bibinfo
		{year} {2019})}\BibitemShut {NoStop}%
	\bibitem [{\citenamefont {Silva}(2014)}]{Silva2014}%
	\BibitemOpen
	\bibfield  {author} {\bibinfo {author} {\bibfnamefont {G.~T.}\ \bibnamefont
			{Silva}},\ }\href {\doibase 10.1121/1.4895691} {\bibfield  {journal}
		{\bibinfo  {journal} {J. Acoust. Soc. Am.}\ }\textbf {\bibinfo {volume}
			{136}},\ \bibinfo {pages} {2405} (\bibinfo {year} {2014})}\BibitemShut
	{NoStop}%
	\bibitem [{\citenamefont {Zhang}\ and\ \citenamefont
		{Marston}(2011{\natexlab{a}})}]{Zhang2011c}%
	\BibitemOpen
	\bibfield  {author} {\bibinfo {author} {\bibfnamefont {L.}~\bibnamefont
			{Zhang}}\ and\ \bibinfo {author} {\bibfnamefont {P.~L.}\ \bibnamefont
			{Marston}},\ }\href@noop {} {\bibfield  {journal} {\bibinfo  {journal} {J.
				Acoust. Soc. Am.}\ }\textbf {\bibinfo {volume} {129}},\ \bibinfo {pages}
		{1679} (\bibinfo {year} {2011}{\natexlab{a}})}\BibitemShut {NoStop}%
	\bibitem [{\citenamefont {Zhang}\ and\ \citenamefont
		{Marston}(2011{\natexlab{b}})}]{Zhang2011a}%
	\BibitemOpen
	\bibfield  {author} {\bibinfo {author} {\bibfnamefont {L.}~\bibnamefont
			{Zhang}}\ and\ \bibinfo {author} {\bibfnamefont {P.~L.}\ \bibnamefont
			{Marston}},\ }\href@noop {} {\bibfield  {journal} {\bibinfo  {journal} {Phys.
				Rev. E}\ }\textbf {\bibinfo {volume} {84}},\ \bibinfo {pages} {065601}
		(\bibinfo {year} {2011}{\natexlab{b}})}\BibitemShut {NoStop}%
	\bibitem [{\citenamefont {Silva}\ \emph {et~al.}(2012)\citenamefont {Silva},
		\citenamefont {Lobo},\ and\ \citenamefont {Mitri}}]{Silva2012}%
	\BibitemOpen
	\bibfield  {author} {\bibinfo {author} {\bibfnamefont {G.~T.}\ \bibnamefont
			{Silva}}, \bibinfo {author} {\bibfnamefont {T.~P.}\ \bibnamefont {Lobo}}, \
		and\ \bibinfo {author} {\bibfnamefont {F.~G.}\ \bibnamefont {Mitri}},\
	}\href@noop {} {\bibfield  {journal} {\bibinfo  {journal} {Europhys. Lett.}\
		}\textbf {\bibinfo {volume} {97}},\ \bibinfo {pages} {54003} (\bibinfo {year}
		{2012})}\BibitemShut {NoStop}%
	\bibitem [{\citenamefont {Mitri}\ \emph {et~al.}(2012)\citenamefont {Mitri},
		\citenamefont {Lobo},\ and\ \citenamefont {Silva}}]{Mitri2012a}%
	\BibitemOpen
	\bibfield  {author} {\bibinfo {author} {\bibfnamefont {F.~G.}\ \bibnamefont
			{Mitri}}, \bibinfo {author} {\bibfnamefont {T.~P.}\ \bibnamefont {Lobo}}, \
		and\ \bibinfo {author} {\bibfnamefont {G.~T.}\ \bibnamefont {Silva}},\
	}\href@noop {} {\bibfield  {journal} {\bibinfo  {journal} {Phys. Rev. E}\
		}\textbf {\bibinfo {volume} {85}},\ \bibinfo {pages} {026602} (\bibinfo
		{year} {2012})}\BibitemShut {NoStop}%
	\bibitem [{\citenamefont {Zhang}\ and\ \citenamefont
		{Marston}(2013)}]{Zhang2013}%
	\BibitemOpen
	\bibfield  {author} {\bibinfo {author} {\bibfnamefont {L.}~\bibnamefont
			{Zhang}}\ and\ \bibinfo {author} {\bibfnamefont {P.~L.}\ \bibnamefont
			{Marston}},\ }\href@noop {} {\bibfield  {journal} {\bibinfo  {journal}
			{Biomed. Opt. Expr.}\ }\textbf {\bibinfo {volume} {4}},\ \bibinfo {pages}
		{1610} (\bibinfo {year} {2013})}\BibitemShut {NoStop}%
	\bibitem [{\citenamefont {Mitri}(2016)}]{Mitri2016}%
	\BibitemOpen
	\bibfield  {author} {\bibinfo {author} {\bibfnamefont {F.~G.}\ \bibnamefont
			{Mitri}},\ }\href@noop {} {\bibfield  {journal} {\bibinfo  {journal}
			{Ultrasonics}\ }\textbf {\bibinfo {volume} {72}},\ \bibinfo {pages} {57}
		(\bibinfo {year} {2016})}\BibitemShut {NoStop}%
	\bibitem [{\citenamefont {Zhang}(2018)}]{Zhang2018}%
	\BibitemOpen
	\bibfield  {author} {\bibinfo {author} {\bibfnamefont {L.}~\bibnamefont
			{Zhang}},\ }\href@noop {} {\bibfield  {journal} {\bibinfo  {journal} {Phys.
				Rev. Applied}\ }\textbf {\bibinfo {volume} {10}},\ \bibinfo {pages} {034039}
		(\bibinfo {year} {2018})}\BibitemShut {NoStop}%
	\bibitem [{\citenamefont {Gong}\ \emph {et~al.}(2019)\citenamefont {Gong},
		\citenamefont {Marston},\ and\ \citenamefont {Li}}]{Gong2019}%
	\BibitemOpen
	\bibfield  {author} {\bibinfo {author} {\bibfnamefont {Z.}~\bibnamefont
			{Gong}}, \bibinfo {author} {\bibfnamefont {P.~L.}\ \bibnamefont {Marston}}, \
		and\ \bibinfo {author} {\bibfnamefont {W.}~\bibnamefont {Li}},\ }\href@noop
	{} {\bibfield  {journal} {\bibinfo  {journal} {Phys. Rev. Appl.}\ }\textbf
		{\bibinfo {volume} {11}},\ \bibinfo {pages} {064022} (\bibinfo {year}
		{2019})}\BibitemShut {NoStop}%
	\bibitem [{\citenamefont {Schwarz}\ \emph {et~al.}(2015)\citenamefont
		{Schwarz}, \citenamefont {Hahn}, \citenamefont {Petit-Pierre},\ and\
		\citenamefont {Dual}}]{Schwarz2015}%
	\BibitemOpen
	\bibfield  {author} {\bibinfo {author} {\bibfnamefont {T.}~\bibnamefont
			{Schwarz}}, \bibinfo {author} {\bibfnamefont {P.}~\bibnamefont {Hahn}},
		\bibinfo {author} {\bibfnamefont {G.}~\bibnamefont {Petit-Pierre}}, \ and\
		\bibinfo {author} {\bibfnamefont {J.}~\bibnamefont {Dual}},\ }\href@noop {}
	{\bibfield  {journal} {\bibinfo  {journal} {Microfluid Nanofluid}\ }\textbf
		{\bibinfo {volume} {18}},\ \bibinfo {pages} {65} (\bibinfo {year}
		{2015})}\BibitemShut {NoStop}%
	\bibitem [{\citenamefont {Wang}\ \emph {et~al.}(2012)\citenamefont {Wang},
		\citenamefont {Castro}, \citenamefont {Hoyos},\ and\ \citenamefont
		{Mallouk}}]{Wang2012}%
	\BibitemOpen
	\bibfield  {author} {\bibinfo {author} {\bibfnamefont {W.}~\bibnamefont
			{Wang}}, \bibinfo {author} {\bibfnamefont {L.~A.}\ \bibnamefont {Castro}},
		\bibinfo {author} {\bibfnamefont {M.}~\bibnamefont {Hoyos}}, \ and\ \bibinfo
		{author} {\bibfnamefont {T.~E.}\ \bibnamefont {Mallouk}},\ }\href@noop {}
	{\bibfield  {journal} {\bibinfo  {journal} {ACS Nano}\ }\textbf {\bibinfo
			{volume} {67}},\ \bibinfo {pages} {6122} (\bibinfo {year}
		{2012})}\BibitemShut {NoStop}%
	\bibitem [{\citenamefont {Wijaya}\ and\ \citenamefont
		{Lim}(2015)}]{Wijaya2015}%
	\BibitemOpen
	\bibfield  {author} {\bibinfo {author} {\bibfnamefont {F.~B.}\ \bibnamefont
			{Wijaya}}\ and\ \bibinfo {author} {\bibfnamefont {K.-M.}\ \bibnamefont
			{Lim}},\ }\href@noop {} {\bibfield  {journal} {\bibinfo  {journal} {Acta
				Acust. united Ac.}\ }\textbf {\bibinfo {volume} {101}},\ \bibinfo {pages}
		{531} (\bibinfo {year} {2015})}\BibitemShut {NoStop}%
	\bibitem [{\citenamefont {Jerome}\ \emph {et~al.}(2019)\citenamefont {Jerome},
		\citenamefont {Ilinskii}, \citenamefont {Zabolotskaya},\ and\ \citenamefont
		{Hamilton}}]{Jerome2019}%
	\BibitemOpen
	\bibfield  {author} {\bibinfo {author} {\bibfnamefont {T.~S.}\ \bibnamefont
			{Jerome}}, \bibinfo {author} {\bibfnamefont {Y.~A.}\ \bibnamefont
			{Ilinskii}}, \bibinfo {author} {\bibfnamefont {E.~A.}\ \bibnamefont
			{Zabolotskaya}}, \ and\ \bibinfo {author} {\bibfnamefont {M.~F.}\
			\bibnamefont {Hamilton}},\ }\href@noop {} {\bibfield  {journal} {\bibinfo
			{journal} {J. Acoust. Soc. Am.}\ }\textbf {\bibinfo {volume} {145}},\
		\bibinfo {pages} {36} (\bibinfo {year} {2019})}\BibitemShut {NoStop}%
	\bibitem [{\citenamefont {Fan}\ \emph {et~al.}(2008)\citenamefont {Fan},
		\citenamefont {Mei}, \citenamefont {Yang},\ and\ \citenamefont
		{Chen}}]{Fan2008}%
	\BibitemOpen
	\bibfield  {author} {\bibinfo {author} {\bibfnamefont {Z.}~\bibnamefont
			{Fan}}, \bibinfo {author} {\bibfnamefont {D.}~\bibnamefont {Mei}}, \bibinfo
		{author} {\bibfnamefont {K.}~\bibnamefont {Yang}}, \ and\ \bibinfo {author}
		{\bibfnamefont {Z.}~\bibnamefont {Chen}},\ }\href@noop {} {\bibfield
		{journal} {\bibinfo  {journal} {J. Acoust. Soc. Am.}\ }\textbf {\bibinfo
			{volume} {124}},\ \bibinfo {pages} {2727} (\bibinfo {year}
		{2008})}\BibitemShut {NoStop}%
	\bibitem [{\citenamefont {Silva}\ \emph {et~al.}(2013)\citenamefont {Silva},
		\citenamefont {Lopes},\ and\ \citenamefont {Mitri}}]{Silva2013}%
	\BibitemOpen
	\bibfield  {author} {\bibinfo {author} {\bibfnamefont {G.~T.}\ \bibnamefont
			{Silva}}, \bibinfo {author} {\bibfnamefont {J.~H.}\ \bibnamefont {Lopes}}, \
		and\ \bibinfo {author} {\bibfnamefont {F.~G.}\ \bibnamefont {Mitri}},\ }\href
	{\doibase 10.1109/TUFFC.2013.2683} {\bibfield  {journal} {\bibinfo  {journal}
			{IEEE Trans. Ultrason. Ferroelectr. Freq. Control}\ }\textbf {\bibinfo
			{volume} {60}},\ \bibinfo {pages} {1207} (\bibinfo {year}
		{2013})}\BibitemShut {NoStop}%
	\bibitem [{\citenamefont {Zhang}\ and\ \citenamefont
		{Marston}(2011{\natexlab{c}})}]{Zhang2011}%
	\BibitemOpen
	\bibfield  {author} {\bibinfo {author} {\bibfnamefont {L.}~\bibnamefont
			{Zhang}}\ and\ \bibinfo {author} {\bibfnamefont {P.~L.}\ \bibnamefont
			{Marston}},\ }\href {\doibase 10.1103/PhysRevE.84.035601} {\bibfield
		{journal} {\bibinfo  {journal} {Phys. Rev. E}\ }\textbf {\bibinfo {volume}
			{84}},\ \bibinfo {pages} {035601} (\bibinfo {year}
		{2011}{\natexlab{c}})}\BibitemShut {NoStop}%
	\bibitem [{\citenamefont {{Le\~ao-Neto}}\ and\ \citenamefont
		{Silva}(2016)}]{Leao-Neto2016}%
	\BibitemOpen
	\bibfield  {author} {\bibinfo {author} {\bibfnamefont {J.~P.}\ \bibnamefont
			{{Le\~ao-Neto}}}\ and\ \bibinfo {author} {\bibfnamefont {G.~T.}\ \bibnamefont
			{Silva}},\ }\href@noop {} {\bibfield  {journal} {\bibinfo  {journal}
			{Ultrasonics}\ }\textbf {\bibinfo {volume} {71}},\ \bibinfo {pages} {1}
		(\bibinfo {year} {2016})}\BibitemShut {NoStop}%
	\bibitem [{\citenamefont {Kong}\ \emph {et~al.}(2012)\citenamefont {Kong},
		\citenamefont {Cui}, \citenamefont {Pan},\ and\ \citenamefont
		{Zhang}}]{Kong2012}%
	\BibitemOpen
	\bibfield  {author} {\bibinfo {author} {\bibfnamefont {D.}~\bibnamefont
			{Kong}}, \bibinfo {author} {\bibfnamefont {Z.}~\bibnamefont {Cui}}, \bibinfo
		{author} {\bibfnamefont {Y.}~\bibnamefont {Pan}}, \ and\ \bibinfo {author}
		{\bibfnamefont {K.}~\bibnamefont {Zhang}},\ }\href@noop {} {\bibfield
		{journal} {\bibinfo  {journal} {Intl. J. Pure Appl. Math.}\ }\textbf
		{\bibinfo {volume} {75}},\ \bibinfo {pages} {455} (\bibinfo {year}
		{2012})}\BibitemShut {NoStop}%
	\bibitem [{\citenamefont {Marzo}\ \emph {et~al.}(2015)\citenamefont {Marzo},
		\citenamefont {Seah}, \citenamefont {Drinkwater}, \citenamefont {Sahoo},
		\citenamefont {Long},\ and\ \citenamefont {Subramanian}}]{Marzo2015}%
	\BibitemOpen
	\bibfield  {author} {\bibinfo {author} {\bibfnamefont {A.}~\bibnamefont
			{Marzo}}, \bibinfo {author} {\bibfnamefont {S.~A.}\ \bibnamefont {Seah}},
		\bibinfo {author} {\bibfnamefont {B.~W.}\ \bibnamefont {Drinkwater}},
		\bibinfo {author} {\bibfnamefont {D.~R.}\ \bibnamefont {Sahoo}}, \bibinfo
		{author} {\bibfnamefont {B.}~\bibnamefont {Long}}, \ and\ \bibinfo {author}
		{\bibfnamefont {S.}~\bibnamefont {Subramanian}},\ }\href {\doibase
		10.1038/ncomms9661} {\bibfield  {journal} {\bibinfo  {journal} {Nat.
				Commun.}\ }\textbf {\bibinfo {volume} {6}},\ \bibinfo {pages} {8661}
		(\bibinfo {year} {2015})}\BibitemShut {NoStop}%
	\bibitem [{\citenamefont {Baudoin}\ \emph {et~al.}(2019)\citenamefont
		{Baudoin}, \citenamefont {Gerbedoen}, \citenamefont {Riaud}, \citenamefont
		{Matar}, \citenamefont {Smagin},\ and\ \citenamefont {Thomas}}]{Baudoin2019}%
	\BibitemOpen
	\bibfield  {author} {\bibinfo {author} {\bibfnamefont {M.}~\bibnamefont
			{Baudoin}}, \bibinfo {author} {\bibfnamefont {J.-C.}\ \bibnamefont
			{Gerbedoen}}, \bibinfo {author} {\bibfnamefont {A.}~\bibnamefont {Riaud}},
		\bibinfo {author} {\bibfnamefont {O.~B.}\ \bibnamefont {Matar}}, \bibinfo
		{author} {\bibfnamefont {N.}~\bibnamefont {Smagin}}, \ and\ \bibinfo {author}
		{\bibfnamefont {J.-L.}\ \bibnamefont {Thomas}},\ }\href@noop {} {\bibfield
		{journal} {\bibinfo  {journal} {Sci. Adv.}\ }\textbf {\bibinfo {volume}
			{5}},\ \bibinfo {pages} {eaav1967} (\bibinfo {year} {2019})}\BibitemShut
	{NoStop}%
	\bibitem [{\citenamefont {Flammer}(2005)}]{Flammer2005}%
	\BibitemOpen
	\bibfield  {author} {\bibinfo {author} {\bibfnamefont {C.}~\bibnamefont
			{Flammer}},\ }\href@noop {} {\emph {\bibinfo {title} {Spheroidal Wave
				Functions}}}\ (\bibinfo  {publisher} {Dover Publications},\ \bibinfo
	{address} {London},\ \bibinfo {year} {2005})\BibitemShut {NoStop}%
	\bibitem [{\citenamefont {Silva}\ and\ \citenamefont
		{Drinkwater}(2018)}]{Silva2018}%
	\BibitemOpen
	\bibfield  {author} {\bibinfo {author} {\bibfnamefont {G.~T.}\ \bibnamefont
			{Silva}}\ and\ \bibinfo {author} {\bibfnamefont {B.~W.}\ \bibnamefont
			{Drinkwater}},\ }\href@noop {} {\bibfield  {journal} {\bibinfo  {journal} {J.
				Acoustic. Soc. Am.}\ }\textbf {\bibinfo {volume} {144}},\ \bibinfo {pages}
		{EL453} (\bibinfo {year} {2018})}\BibitemShut {NoStop}%
	\bibitem [{\citenamefont {Silva}(2011{\natexlab{a}})}]{Silva2011}%
	\BibitemOpen
	\bibfield  {author} {\bibinfo {author} {\bibfnamefont {G.~T.}\ \bibnamefont
			{Silva}},\ }\href@noop {} {\bibfield  {journal} {\bibinfo  {journal} {J.
				Acoust. Soc. Am.}\ }\textbf {\bibinfo {volume} {130}},\ \bibinfo {pages}
		{3541} (\bibinfo {year} {2011}{\natexlab{a}})}\BibitemShut {NoStop}%
	\bibitem [{\citenamefont {Lopes}\ \emph {et~al.}(2016)\citenamefont {Lopes},
		\citenamefont {Azarpeyvand},\ and\ \citenamefont {Silva}}]{Lopes2016}%
	\BibitemOpen
	\bibfield  {author} {\bibinfo {author} {\bibfnamefont {J.~H.}\ \bibnamefont
			{Lopes}}, \bibinfo {author} {\bibfnamefont {M.}~\bibnamefont {Azarpeyvand}},
		\ and\ \bibinfo {author} {\bibfnamefont {G.~T.}\ \bibnamefont {Silva}},\
	}\href {\doibase 10.1109/TUFFC.2015.2494693} {\bibfield  {journal} {\bibinfo
			{journal} {IEEE Trans. Ultrason. Ferroelectr. Freq. Control}\ }\textbf
		{\bibinfo {volume} {63}},\ \bibinfo {pages} {186} (\bibinfo {year}
		{2016})}\BibitemShut {NoStop}%
	\bibitem [{\citenamefont {Gong}\ and\ \citenamefont
		{Marston}(2017)}]{Gong2017}%
	\BibitemOpen
	\bibfield  {author} {\bibinfo {author} {\bibfnamefont {Z.}~\bibnamefont
			{Gong}}\ and\ \bibinfo {author} {\bibfnamefont {P.~L.}\ \bibnamefont
			{Marston}},\ }\href@noop {} {\bibfield  {journal} {\bibinfo  {journal} {J.
				Acoust. Soc. Am.}\ }\textbf {\bibinfo {volume} {141}},\ \bibinfo {pages}
		{EL574} (\bibinfo {year} {2017})}\BibitemShut {NoStop}%
	\bibitem [{\citenamefont {Mitri}\ and\ \citenamefont
		{Silva}(2014)}]{Mitri2014}%
	\BibitemOpen
	\bibfield  {author} {\bibinfo {author} {\bibfnamefont {F.~G.}\ \bibnamefont
			{Mitri}}\ and\ \bibinfo {author} {\bibfnamefont {G.~T.}\ \bibnamefont
			{Silva}},\ }\href {\doibase 10.1103/PhysRevE.90.053204} {\bibfield  {journal}
		{\bibinfo  {journal} {Phys. Rev. E}\ }\textbf {\bibinfo {volume} {90}},\
		\bibinfo {pages} {053204} (\bibinfo {year} {2014})}\BibitemShut {NoStop}%
	\bibitem [{\citenamefont {Silva}(2011{\natexlab{b}})}]{Silva2011a}%
	\BibitemOpen
	\bibfield  {author} {\bibinfo {author} {\bibfnamefont {G.~T.}\ \bibnamefont
			{Silva}},\ }\href@noop {} {\bibfield  {journal} {\bibinfo  {journal} {IEEE
				Trans. Ultrason. Ferroelectr. Freq. Control}\ }\textbf {\bibinfo {volume}
			{58}},\ \bibinfo {pages} {298} (\bibinfo {year}
		{2011}{\natexlab{b}})}\BibitemShut {NoStop}%
	\bibitem [{\citenamefont {Mitri}\ and\ \citenamefont
		{Silva}(2011)}]{Mitri2011}%
	\BibitemOpen
	\bibfield  {author} {\bibinfo {author} {\bibfnamefont {F.~G.}\ \bibnamefont
			{Mitri}}\ and\ \bibinfo {author} {\bibfnamefont {G.~T.}\ \bibnamefont
			{Silva}},\ }\href@noop {} {\bibfield  {journal} {\bibinfo  {journal} {Wave
				Motion}\ }\textbf {\bibinfo {volume} {46}},\ \bibinfo {pages} {392} (\bibinfo
		{year} {2011})}\BibitemShut {NoStop}%
	\bibitem [{\citenamefont {Silva}\ \emph {et~al.}(2015)\citenamefont {Silva},
		\citenamefont {Baggio}, \citenamefont {Lopes},\ and\ \citenamefont
		{Mitri}}]{Silva2015a}%
	\BibitemOpen
	\bibfield  {author} {\bibinfo {author} {\bibfnamefont {G.~T.}\ \bibnamefont
			{Silva}}, \bibinfo {author} {\bibfnamefont {A.~L.}\ \bibnamefont {Baggio}},
		\bibinfo {author} {\bibfnamefont {J.~H.}\ \bibnamefont {Lopes}}, \ and\
		\bibinfo {author} {\bibfnamefont {F.~G.}\ \bibnamefont {Mitri}},\ }\href@noop
	{} {\bibfield  {journal} {\bibinfo  {journal} {IEEE Trans. Ultrason.
				Ferroelectr. Freq. Control}\ }\textbf {\bibinfo {volume} {62}},\ \bibinfo
		{pages} {576} (\bibinfo {year} {2015})}\BibitemShut {NoStop}%
	\bibitem [{\citenamefont {Mishchenko}\ \emph {et~al.}(2002)\citenamefont
		{Mishchenko}, \citenamefont {Travis},\ and\ \citenamefont
		{Lacis}}]{Mishchenko2002}%
	\BibitemOpen
	\bibfield  {author} {\bibinfo {author} {\bibfnamefont {M.~I.}\ \bibnamefont
			{Mishchenko}}, \bibinfo {author} {\bibfnamefont {L.~D.}\ \bibnamefont
			{Travis}}, \ and\ \bibinfo {author} {\bibfnamefont {A.~A.}\ \bibnamefont
			{Lacis}},\ }\href@noop {} {\emph {\bibinfo {title} {Scattering, Absorption,
				and Emission of Light by Small Particles}}}\ (\bibinfo  {publisher}
	{Cambridge University Press},\ \bibinfo {address} {Cambridge},\ \bibinfo
	{year} {2002})\BibitemShut {NoStop}%
	\bibitem [{\citenamefont {{Wolfram Research{,} Inc.}}(2019)}]{Mathematica}%
	\BibitemOpen
	\bibfield  {author} {\bibinfo {author} {\bibnamefont {{Wolfram Research{,}
					Inc.}}},\ }\href@noop {} {\enquote {\bibinfo {title} {Mathematica, {V}ersion
				10.0},}\ }\bibinfo {howpublished} {https://www.wolfram.com/mathematica}
	(\bibinfo {year} {2019}),\ \bibinfo {note} {{Champaign, IL}}\BibitemShut
	{NoStop}%
	\bibitem [{\citenamefont {Lide}(2004)}]{Lide2004}%
	\BibitemOpen
	\bibfield  {author} {\bibinfo {author} {\bibfnamefont {D.~R.}\ \bibnamefont
			{Lide}},\ }\href@noop {} {\emph {\bibinfo {title} {CRC Handbook of Chemistry
				and Physics}}},\ \bibinfo {edition} {84th}\ ed.\ (\bibinfo  {publisher} {CRC
		Press},\ \bibinfo {address} {Boca Raton, FL},\ \bibinfo {year}
	{2004})\BibitemShut {NoStop}%
	\bibitem [{\citenamefont {Lee}\ and\ \citenamefont {Wang}(1989)}]{Lee1989}%
	\BibitemOpen
	\bibfield  {author} {\bibinfo {author} {\bibfnamefont {C.~P.}\ \bibnamefont
			{Lee}}\ and\ \bibinfo {author} {\bibfnamefont {T.~G.}\ \bibnamefont {Wang}},\
	}\href@noop {} {\bibfield  {journal} {\bibinfo  {journal} {J. Acoust. Soc.
				Am.}\ }\textbf {\bibinfo {volume} {85}},\ \bibinfo {pages} {1081} (\bibinfo
		{year} {1989})}\BibitemShut {NoStop}%
	\bibitem [{\citenamefont {Burke}(1966)}]{Burke1966}%
	\BibitemOpen
	\bibfield  {author} {\bibinfo {author} {\bibfnamefont {J.~E.}\ \bibnamefont
			{Burke}},\ }\href@noop {} {\bibfield  {journal} {\bibinfo  {journal} {Stud.
				Appl. Math.}\ }\textbf {\bibinfo {volume} {45}},\ \bibinfo {pages} {425}
		(\bibinfo {year} {1966})}\BibitemShut {NoStop}%
\end{thebibliography}
%

\end{document}